\documentclass[12pt,prd,tightenlines,nofootinbib,showpacs,showkeys]{revtex4}
\newcommand{\be}{\begin{equation}}
\newcommand{\ee}{\end{equation}}
\newcommand{\bdis}{\begin{displaymath}}
\newcommand{\edis}{\end{displaymath}}
\newcommand{\bga}{\begin{equation}\begin{gathered}}
\newcommand{\ega}{\end{gathered}\end{equation}}
\usepackage{bm}
\usepackage{graphics}
\usepackage{rotating}
\usepackage{epsfig}
\usepackage{amsmath}
\usepackage{amsfonts}

\begin{document}

\title{
\begin{flushright}
{\rm\normalsize SSU-HEP-16/09}
\end{flushright}
Lamb shift in muonic ions of lithium, beryllium and boron}
\author{\firstname{A.A.} \surname{Krutov}}
\author{\firstname{A.P.} \surname{Martynenko}}
\author{\firstname{F.A.} \surname{Martynenko}}
\author{\firstname{O.S.} \surname{Sukhorukova}}
\affiliation{Samara University, 443086, Moskovskoe shosse 34, Samara, Russia}

\begin{abstract}
We present a precise calculation of the Lamb shift $(2P_{1/2}-2S_{1/2})$ in muonic ions
$(\mu ^6_3Li)^{2+},~(\mu ^7_3Li)^{2+}$, $(\mu ^9_4Be)^{3+},~(\mu ^{10}_4Be)^{3+}$,
$(\mu ^{10}_5B)^{4+},~(\mu ^{11}_5B)^{4+}$. The contributions of orders  $\alpha^3\div\alpha^6$
to the vacuum polarization, nuclear structure and recoil, relativistic effects
are taken into account. Our numerical results are consistent with previous calculations
and improve them due to account of new corrections.
The obtained results can be used for the comparison with future experimental data, and extraction
more accurate values of nuclear charge radii.
\end{abstract}



\maketitle

\section{Introduction}

In recent years a special interest in the physics of elementary particles is related with muons.
Experimental study of the muon anomalous magnetic moment revealed a certain discrepancy between theoretical
and experimental results.
The subsequent in 2010 year measurement of the Lamb shift in muonic hydrogen has led to another problem,
called the proton charge radius puzzle \cite{CREMA,CREMA1,CREMA2,CREMA3}.
After a new measurement of the Lamb shift \cite{CREMA4} in muonic deuterium it became clear that there is
a discrepancy in the values of the charge radius of the proton and deuteron determined
by electronic and muonic atoms.
This may mean that the muons are playing an important role in subatomic physics, which is not fully understood.
The experimental CREMA collaboration program includes other muon atoms, especially muonic helium ions \cite{CREMA5},
but it apparently can be extended to the study of other light muonic atoms. The transition energy $2P\to 2S$
in light muonic atoms can be precisely measured by laser spectroscopy as in muonic hydrogen.
Therefore, additional theoretical study of muon bound states and a calculation of their energy levels,
along with experimental investigations, can contribute to a better understanding of the essence of the problem.

The interest in muonic ions $(\mu Li)^{2+}$, $(\mu Be)^{3+}$, $(\mu B)^{4+}$ is also connected to the fact that,
as has been established in this case,
there is a strong cancelation of two main contributions to the one-loop vacuum polarization
and the structure of the nucleus \cite{Drake1,Drake2}. As a result, the Lamb shift value lies
in a wide range of wavelengths $150\div 1100$ nm from ultraviolet to infrared region of the spectrum,
making it possible for its study of laser spectroscopy methods. The measurement of transition frequencies
gives an opportunity to obtain more exact information about nuclear size and structure.
Another important conclusion arising out of this cancelation is that a more significant role
than usual, beginning to play the contributions of higher order in $\alpha$, as well as contributions
containing large degrees of nuclear charge Z. Their methodical analysis is very important to increase the accuracy
of calculation of the Lamb shift \cite{m1999,m2005}. As usual, the most important corrections in the Lamb shift
are the corrections to the vacuum polarization, the nuclear structure and recoil, as well as complex
combination corrections to the vacuum polarization and relativism, which we explore in this paper.

Fundamentals of calculating of energy spectra for light muonic atoms were formulated many years ago
in relativistic approach based on the Dirac equation, and in nonrelativistic Schr\"odinger method
in \cite{BR1,BR2,BR3,friar,KP1,EGS} (see other references in review articles \cite{BR3,EGS}).
After recent experiments of the CREMA collaboration
in 2010 year there were many works devoted to the muonic atoms, in order to overcome the arisen difference
in the magnitude of the charge radius of the proton \cite{borie,udj,cec,kik,friar2013,cec2014,kkis,miller,ar,krauth}
(see other references in \cite{CREMA2}).
They carried out a new analysis of the main contribution to the Lamb shift and hyperfine structure of the
spectrum and the various corrections that have quite significant numerical value.
There was also analyzed a number of relatively subtle effects in the fine and hyperfine structure,
which however did not lead to any considerable change in the results (see, for example, \cite{efmm,fmms,vp,kou}).
The aim of the present work is to extend our previous calculations of the Lamb shift in muonic helium ions
\cite{jetp2015} on other muonic ions such as muonic lithium, muonic beryllium and muonic boron.
We consistently calculate the contributions of orders $\alpha^3\div \alpha^6$
within the framework of the quasipotential method in quantum electrodynamics \cite{m1,m2,m3,m4}.
It is important to know also the hyperfine splitting of levels for the evaluation of observed transition frequencies \cite{Drake2}.
As in previous works we take modern numerical values of fundamental physical constants from \cite{MT,marinova}.
Since the corrections to the structure of the nucleus play a key role further, let us write explicitly
used values of the nuclear charge radii: $r(_3^6Li)=(2.5890\pm 0.0390)$ fm, $r(_3^7Li)=(2.4440\pm 0.0420)$ fm,
$r(_4^9Be)=(2.5190\pm 0.0120)$ fm,
$r(_4^{10}Be)=(2.3550\pm 0.0170)$ fm, $r(_5^{10}B)=(2.4277\pm 0.0499)$ fm, $r(_5^{11}B)=(2.4060\pm 0.0294)$ fm.

\section{Effects of vacuum polarization in the one-photon interaction}

Let us begin by recalling the basic assumptions of the quasipotential approach in the calculation of the
Lamb shift. Muonic ion is described by the Schr\"odinger equation with the Breit Hamiltonian \cite{t4}:
\begin{equation}
\label{eq:1}
H_B=\frac{{\bf p}^2}{2\mu}-\frac{Z\alpha}{r}-\frac{{\bf
p}^4}{8m_1^3}- \frac{{\bf p}^4}{8m_2^3}+\frac{\pi
Z\alpha}{2}\left(\frac{1}{m_1^2}+
\frac{\delta_I}{m_2^2}\right)\delta({\bf r})-
\end{equation}
\begin{displaymath}
-\frac{Z\alpha}{2m_1m_2r}\left({\bf p}^2+\frac{{\bf r}({\bf
rp}){\bf p}}
{r^2}\right)+\frac{Z\alpha}{r^3}\left(\frac{1}{4m_1^2}+\frac{1}{2m_1m_2}\right)
({\bf L}{\mathstrut\bm\sigma}_1)=H_0+\Delta V^B,
\end{displaymath}
where $H_0={\bf p}^2/2\mu-Z\alpha/r$, $m_1$, $m_2$ are the muon
and nucleus masses, $\mu=m_1m_2/(m_1+m_2)$, $\delta_I=1$ for nucleus with half-integer spin,
$\delta_I=0$ for nucleus with integer spin. The exact solution of the Schr\"odinger equation with the
Hamiltonian $H_0$ is then used in the calculation of the shifts of $2S$ and $2$P energy levels
by perturbation theory.

As is well known the basic contribution to the Lamb shift in muonic atoms is determined by the effect
of electron vacuum polarization (VP) in $1\gamma$-interaction. The potential of particle interaction
corresponding to this effect has the form:
\begin{equation}
\label{eq:2}
V^C_{vp}(r)=\frac{\alpha}{3\pi}\int_1^\infty d\xi\rho(\xi) \left(-\frac{Z\alpha}{r}e^{-2m_e\xi r}\right),~~~
\rho(\xi)=\frac{\sqrt{\xi^2-1}(2\xi^2+1)}{\xi^4}.
\end{equation}
It gives the shift of energy levels for $2S$ and $2P$ states which can be presented in analytical form
($k_1=2m_e/W$, $W=\mu Z\alpha$):
\begin{equation}
\label{eq:3}
\Delta E_{vp}(2S)=-\frac{\mu(Z\alpha)^2\alpha}{6\pi}\int_1^\infty
\rho(\xi)d\xi\int_0^\infty x dx\left(1-\frac{x}{2}\right)^2e^{-x\left(1+
\frac{2m_e\xi}{W}\right)}=
\end{equation}
\begin{displaymath}
\frac{1}{12 \left(1-k_1^2\right)^{5/2}}
\Bigl[\sqrt{1-k_1^2} \left(-168 k_1^6+272 k_1^4-49 k_1^2+6 \pi  \left(k_1^2-1\right)^2 \left(14
k_1^2+3\right) k_1-28\right)+
\end{displaymath}
\begin{displaymath}
+3 \left(56 k_1^8-128 k_1^6+75 k_1^4+10 k_1^2-4\right) \ln \left(\frac{1-\sqrt{1-k_1^2}}{k_1}\right)\Bigr],
\end{displaymath}
\begin{equation}
\label{eq:4}
\Delta E_{vp}(2P)=-\frac{\mu(Z\alpha)^2\alpha}{72\pi}\int_1^\infty
\rho(\xi)d\xi\int_0^\infty x^3 dx e^{-x\left(1+
\frac{2m_e\xi}{W}\right)}=
\end{equation}
\begin{displaymath}
=\frac{1}{\left(1-k_1^2\right)^{5/2}}\Bigl[\sqrt{1-k_1^2} \left(-120 k_1^6+184 k_1^4-23 k_1^2+6 \pi
\left(k_1^2-1\right)^2 \left(10k_1^2+3\right) k_1-32\right)
\end{displaymath}
\begin{displaymath}
+3 \left(40 k_1^8-88 k_1^6+45 k_1^4+10 k_1^2-4\right) \ln
\left(\frac{1-\sqrt{1-k_1^2}}{k_1}\right)\Bigr].
\end{displaymath}
Expressions \eqref{eq:3} and \eqref{eq:4} give the following numerical values to the Lamb shift in the muonic ions:
\begin{equation}
\label{eq:5}
\Delta E_{vp}(2P-2S)=\begin{cases}
^6_3Li:4664.95~meV,&^7_3Li:4682.38~meV\\
^9_4Be:9255.79~meV,&^{10}_4Be:9270.74~meV\\
^{10}_5B:15356.42~meV,&^{11}_5B:15375.55~meV
\end{cases}.
\end{equation}
We retain two significant figures after the decimal point in all obtained expressions.
The order of contributions \eqref{eq:3} and \eqref{eq:4} is clearly extracted in front
of integrals. For the calculation of muon VP contribution we use again
\eqref{eq:3} and \eqref{eq:4} changing $m_e\to m_\mu$. Corresponding numerical values
which have the order $\alpha(Z\alpha)^4$ are included in Tables~\ref{tb1},\ref{tb2},\ref{tb3}.

The two-loop vacuum polarization effects in the one-photon interaction can be divided into two
parts: loop-after-loop correction (vp-vp) and two-loop vacuum polarization operator correction which we denote
further as the "2-loop vp" correction. The potential of loop-after-loop VP effect has the form \cite{m2005,jetp2015}:
\begin{equation}
\label{eq:6}
V^C_{vp-vp}(r)=\frac{\alpha^2}{9\pi^2}
\int_1^\infty\rho(\xi)d\xi\int_1^\infty\rho(\eta)d\eta\left(-\frac{Z\alpha}{r}
\right)\frac{1}{(\xi^2-\eta^2)}\left(\xi^2e^{-2m_e\xi
r}-\eta^2e^{-2m_e\eta r} \right).
\end{equation}
Calculating the matrix elements of the potential \eqref{eq:5} in the first order perturbation
theory, we find the contribution to the Lamb shift of order $\alpha^2(Z\alpha)^2$:
\begin{equation}
\label{eq:7}
\Delta E_{vp-vp}(2P-2S)=\begin{cases}
^6_3Li:14.20~meV,&^7_3Li:14.28~meV\\
^9_4Be:33.79~meV,&^{10}_4Be:34.06~meV\\
^{10}_5B:64.40~meV,&^{11}_5B:64.52~meV
\end{cases}.
\end{equation}

There is another correction to the potential which is determined by the amplitude with two
sequential electron and muon loops:
\begin{equation}
\label{eq:8}
\Delta V_{vp-mvp}(r)=-\frac{4(Z\alpha)\alpha^2}{45\pi^2m_1^2}\int_1^\infty
\rho(\xi)d\xi\left[\pi\delta({\bf r})-\frac{m_e^2\xi^2}{r}e^{-2m_e\xi r}\right].
\end{equation}
It gives the correction of order $\alpha^2(Z\alpha)^4$ to the Lamb shift $(2P-2S)$ which is included
in  Tables~\ref{tb1},\ref{tb2},\ref{tb3}.

Two-loop polarization operator contribution to the potential can be presented in the form similar to \eqref{eq:2}
with more complicated spectral function $f(v)$ \cite{ks}:
\begin{equation}
\label{eq:9}
\Delta V_{2-loop~vp}^C=-\frac{2}{3}\frac{Z\alpha}{r}
\left(\frac{\alpha}{\pi}\right)^2\int_0^1\frac{f(v)dv}{(1-v^2)}
e^{-\frac{2m_er}{\sqrt{1-v^2}}},
\end{equation}
\begin{equation}
\label{eq:10}
f(v)=v\Bigl\{(3-v^2)(1+v^2)\left[Li_2\left(-\frac{1-v}{1+v}\right)+2Li_2
\left(\frac{1-v}{1+v}\right)+\frac{3}{2}\ln\frac{1+v}{1-v}\ln\frac{1+v}{2}-
\ln\frac{1+v}{1-v}\ln v\right]
\end{equation}
\begin{displaymath}
+\left[\frac{11}{16}(3-v^2)(1+v^2)+\frac{v^4}{4}\right]\ln\frac{1+v}{1-v}+
\left[\frac{3}{2}v(3-v^2)\ln\frac{1-v^2}{4}-2v(3-v^2)\ln v\right]+
\frac{3}{8}v(5-3v^2)\Bigr\},
\end{displaymath}
where $Li_2(z)$ is the Euler dilogarithm. The potential $\Delta V^C_{2-loop~vp}(r)$ gives the
contribution to the Lamb shift $(2P-2S)$ of order $\alpha^2(Z\alpha)^2$:
\begin{equation}
\label{eq:11}
\Delta E_{2-loop~vp}(2P-2S)=\begin{cases}
^6_3Li:18.21~meV,&^7_3Li:18.26~meV\\
^9_4Be:31.66~meV,&^{10}_4Be:31.69~meV\\
^{10}_5B:47.54~meV,&^{11}_5B:47.57~meV
\end{cases}.
\end{equation}

Numerical value of corrections \eqref{eq:7} and \eqref{eq:11} show that at necessary
level of accuracy we should calculate three-loop VP contributions in one-photon interaction.
One part of three-loop VP effects with successive loops in the scattering amplitude
(loop-after-loop-after-loop, two-loop-after-loop)
can be derived as potential \eqref{eq:6}. Corresponding contributions to the potential and
the Lamb shift $(2P-2S)$ are the following:
\begin{equation}
\label{eq:12}
V^C_{vp-vp-vp}(r)=-\frac{Z\alpha}{r}\frac{\alpha^3}{(3\pi)^3}\int_1^\infty
\rho(\xi)d\xi\int_1^\infty\rho(\eta
d\eta\int_1^\infty\rho(\zeta)d\zeta\times
\end{equation}
\begin{displaymath}
\times\left[e^{-2m_e\zeta
r}\frac{\zeta^4}{(\xi^2-\zeta^2)(\eta^2-\zeta^2)} +e^{-2m_e\xi
r}\frac{\xi^4}{(\zeta^2-\xi^2)(\eta^2-\xi^2)}+ e^{-2m_e\eta
r}\frac{\eta^4}{(\xi^2-\eta^2)(\zeta^2-\eta^2)}\right],
\end{displaymath}
\begin{equation}
\label{eq:13}
V^C_{vp-2-loop~vp}=-\frac{4\mu\alpha^3(Z\alpha)}{9\pi^3}\int_1^\infty
\rho(\xi)d\xi\int_1^\infty\frac{f(\eta)d\eta}{\eta}\frac{1}{r(\eta^2-\xi^2)}
\left(\eta^2e^{-2m_e\eta r}-\xi^2e^{-2m_e\xi r}\right),
\end{equation}
\begin{equation}
\label{eq:14}
\Delta E_{vp-vp-vp}(2P-2S)=\begin{cases}
^6_3Li:0.04~meV,&^7_3Li:0.04~meV\\
^9_4Be:0.11~meV,&^{10}_4Be:0.11~meV\\
^{10}_5B:0.20~meV,&^{11}_5B:0.24~meV,
\end{cases}
\end{equation}
\begin{equation}
\label{eq:15}
\Delta E_{vp-2-loop~vp}(2P-2S)=\begin{cases}
^6_3Li:0.12~meV,&^7_3Li:0.12~meV\\
^9_4Be:0.27~meV,&^{10}_4Be:0.27~meV\\
^{10}_5B:0.48~meV,&^{11}_5B:0.48~meV.
\end{cases}.
\end{equation}

Another part of the diagrams corresponds to the three-loop corrections to the polarization
operator. They were first calculated for the $(2P-2S)$ Lamb shift in muonic hydrogen in \cite{KN1,KN2}.
An estimate of their contribution to the Lamb shift is included in Tables~\ref{tb1}-\ref{tb3}.
Finally, there exists another one-loop vacuum polarization correction of order $\alpha(Z\alpha)^4$
in the Lamb shift known as the Wichmann-Kroll correction \cite{WK,MPS}. Its calculation was discussed
repeatedly in \cite{EGS,jetp2015}, so we restrict ourselves here by including numerical results in the final Tables
as well as the whole light-by-light contribution (see detailed calculation in \cite{sgkjetpl}).
Almost all of the corrections presented in this section are written in the integral form,
and are therefore specific character for each muon atom.
The presented numerical values provide important information about the change in the value
of corrections in different muonic ions.

\section{Relativistic corrections with the account of vacuum polarization effects}

The electron vacuum polarization effects modify not only the Coulomb potential,
but also all other terms of the Breit Hamiltonian. Appropriate potentials, which take into account
relativistic effects and vacuum polarization effects were built in \cite{m2005,KP1,udjpra2011,m2007}:
\begin{equation}
\label{eq:16}
\Delta V^B_{vp}(r)=\frac{\alpha}{3\pi}\int_1^\infty\rho(\xi)d\xi\sum_{i=1}^4
\Delta V_{i,vp}^B(r),
\end{equation}
\begin{equation}
\label{eq:17}
\Delta V_{1,vp}^B=\frac{Z\alpha}{8}\left(\frac{1}{m_1^2}+\frac{\delta_I}{m_2^2}\right)
\left[4\pi\delta({\bf r})-\frac{4m_e^2\xi^2}{r}e^{-2m_e\xi r}\right],
\end{equation}
\begin{equation}
\label{eq:18}
\Delta V_{2,vp}^B=-\frac{Z\alpha m_e^2\xi^2}{m_1m_2r}e^{-2m_e\xi
r}(1- m_e\xi r),
\end{equation}
\begin{equation}
\label{eq:19}
\Delta V_{3,vp}^B=-\frac{Z\alpha}{2m_1m_2}p_i\frac{e^{-2m_e\xi r}}{r}
\left[\delta_{ij}+\frac{r_ir_j}{r^2}(1+2m_e\xi r)\right]p_j,
\end{equation}
\begin{equation}
\label{eq:20}
\Delta V_{4,vp}^B=\frac{Z\alpha}{r^3}\left(\frac{1}{4m_1^2}+\frac{1}{2m_1m_2}
\right)e^{-2m_e\xi r}(1+2m_e\xi r)({\bf L}{\mathstrut\bm\sigma}_1).
\end{equation}
An averaging of these terms gives the corrections of order $\alpha(Z\alpha)^4$ to
the Lamb shift $(2P-2S)$:
\begin{equation}
\label{eq:21}
\Delta E^{B}_{1,vp}(2P-2S)=\begin{cases}
^6_3Li:-5.55~meV,&^7_3Li:-5.60~meV\\
^9_4Be:-19.94~meV,&^{10}_4Be:-20.02~meV\\
^{10}_5B:-52.76~meV,&^{11}_5B:-52.94~meV
\end{cases},
\end{equation}
\begin{equation}
\label{eq:22}
\Delta E^{B}_{2,vp}(2P-2S)=\begin{cases}
^6_3Li:0.04~meV,&^7_3Li:0.04~meV\\
^9_4Be:0.90~meV,&^{10}_4Be:0.08~meV\\
^{10}_5B:0.20~meV,&^{11}_5B:0.18~meV
\end{cases},
\end{equation}
\begin{equation}
\label{eq:23}
\Delta E^{B}_{3,vp}(2P-2S)=\begin{cases}
^6_3Li:0.10~meV,&^7_3Li:0.09~meV\\
^9_4Be:0.27~meV,&^{10}_4Be:0.25~meV\\
^{10}_5B:0.68~meV,&^{11}_5B:0.62~meV
\end{cases},
\end{equation}
\begin{equation}
\label{eq:24}
\Delta E^{B}_{4,vp}(2P-2S)=\begin{cases}
^6_3Li:-0.66~meV,&^7_3Li:-0.66~meV\\
^9_4Be:-2.65~meV,&^{10}_4Be:-2.65~meV\\
^{10}_5B:-7.58~meV,&^{11}_5B:-7.59~meV
\end{cases}.
\end{equation}
The sum of corrections \eqref{eq:21}-\eqref{eq:24} is included in Tables~\ref{tb1},\ref{tb2},\ref{tb3}.
The next step to refine the results of the Lamb shift calculation is associated with the two-loop corrections
to the polarization of the vacuum in the Breit Hamiltonian. So, for example, two-loop analogue of expression
\eqref{eq:17} is equal to
\begin{equation}
\label{eq:25}
\Delta V_{2-loop~vp}^B(r)=\frac{\alpha^2(Z\alpha)}{12\pi^2}\left(\frac{1}
{m_1^2}+\frac{1}{m_2^2}\right)\int_0^1\frac{f(v)dv}{1-v^2}\left[4\pi
\delta({\bf r})-\frac{4m_e^2}{(1-v^2)r}e^{-\frac{2m_er}{\sqrt{1-v^2}}}\right].
\end{equation}
Corresponding correction to the $(2P-2S)$ shift is on the limit of the accuracy of our calculations.
The contribution of other two-loop corrections to the Breit potential can be roughly estimated
in the energy spectrum  at 10 $\%$ (see summary two-loop result in Tables~\ref{tb1}-\ref{tb3}).

\begin{figure}[htbp]
\centering
\includegraphics[scale=0.7]{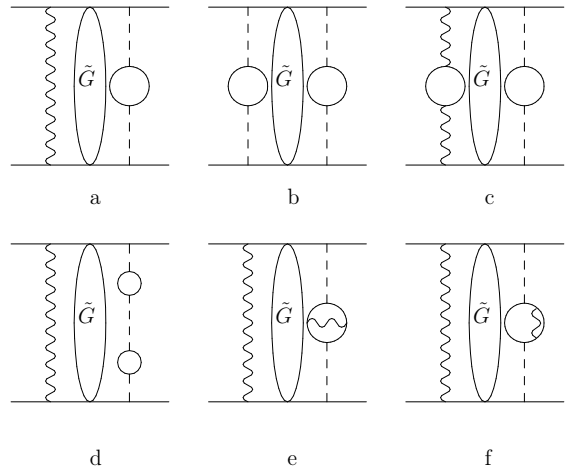}
\caption{Effects of one-loop and two-loop vacuum polarization
in the second order perturbation theory (sopt). Dashed line
shows the Coulomb photon. $\tilde G$ is the reduced Coulomb Green
function (33). Wave line shows terms of the Breit potential.}
\label{fig1}
\end{figure}

In the second order perturbation theory (sopt) there are one-loop and
two-loop electron vacuum polarization contributions of orders $\alpha^2(Z\alpha)^2$
and $\alpha(Z\alpha)^4$. To better understand the structure of these contributions, we present
them on diagrams in Fig.\ref{fig1}. The general expression for corrections has the form:
\begin{equation}
\label{eq:26}
\Delta E_{sopt}^{vp}=<\psi|\Delta V^C_{vp}\tilde
G\Delta V^C_{vp}|\psi>+ 2<\psi|\Delta V^B\tilde G\Delta V^C_{vp}|\psi>,
\end{equation}
where $\tilde G$ is the reduced Coulomb Green's function (RCGF). For the calculation
of the Lamb shift contributions we use a representation of the RCGF for $2S-$ and $2P-$ states
obtained in \cite{hameka} (see exact expressions for $\tilde G_{2S}$, $\tilde G_{2P}$, $g_{2S}$ and
$g_{2P}$ in \cite{jetp2015}). In the case of the two-loop corrections shown in Fig.\ref{fig1}(c),
we get the integral expressions for $2S$ and $2P$ states
\begin{equation}
\label{eq:27}
\Delta E^{vp,vp}_{sopt}(2S)=-\frac{\mu\alpha^2(Z\alpha)^2}{72\pi^2}
\int_1^\infty\rho(\xi)d\xi\int_1^\infty\rho(\eta)d\eta\times
\end{equation}
\begin{displaymath}
\times\int_0^\infty\left(1-\frac{x}{2}\right)e^{-x\left(1-\frac{2m_e\xi}{W}
\right)}dx\int_0^\infty\left(1-\frac{x'}{2}\right)
e^{-x'\left(1-\frac{2m_e\eta}{W}\right)}dx'g_{2S}(x,x'),
\end{displaymath}
\begin{equation}
\label{eq:28}
\Delta E^{vp,vp}_{sopt}(2P)=-\frac{\mu\alpha^2(Z\alpha)^2}{7776\pi^2}\int_1^\infty
\rho(\xi)d\xi\int_1^\infty\rho(\eta)d\eta\times
\end{equation}
\begin{displaymath}
\times\int_0^\infty e^{-x\left(1+\frac{2m_e\xi}{W}\right)}dx
\int_0^\infty e^{-x'\left(1+\frac{2m_e\eta}{W}\right)}
dx'g_{2P}(x,x'),
\end{displaymath}
which then give the following numerical results for the Lamb shift:
\begin{equation}
\label{eq:29}
\Delta E^{vp,vp}_{sopt}(2P-2S)=\begin{cases}
^6_3Li:5.63~meV,&^7_3Li:5.66~meV\\
^9_4Be:12.70~meV,&^{10}_4Be:12.73~meV\\
^{10}_5B:23.39~meV,&^{11}_5B:23.43~meV.
\end{cases}.
\end{equation}
The relations \eqref{eq:27}-\eqref{eq:29} show a sequence of steps in the calculation of the Lamb shift.
Note also that all the integrals over the coordinates of the particles are calculated analytically.
Another contribution, corresponding to the amplitude in Fig.\ref{fig1}(c), is obtained by changing
the perturbation potential with electron vacuum polarization to potential with muon vacuum polarization.
The order of this correction is increased by an additional factor $\alpha^2$.

The second term in \eqref{eq:26} has the similar structure (see Fig.\ref{fig1}(b)). For its evaluation we can
use a number of intermediate algebraic transformation. We show them in the example of one part
of the Breit potential, proportional to ${\bf p}^4/(2\mu)^2$:
\begin{equation}
\label{eq:30}
\bigl <\psi\bigl|\frac{{\bf p}^4}{(2\mu)^2}{\sum}'_m\frac{|\psi_m><\psi_m|}{E_2-E_m}
\Delta V^C_{vp}\bigr|\psi\bigr>=\bigl<\psi\bigl|(E_2+\frac{Z\alpha}{r})(\hat H_0+
\frac{Z\alpha}{r}){\sum}'_m\frac{|\psi_m><\psi_m|}{E_2-E_m}\Delta V_{vp}^C \bigr|\psi\bigr>=
\end{equation}
\begin{displaymath}
=\bigl<\psi\bigl|\bigl(E_2+\frac{Z\alpha}{r}\bigr)^2\tilde G\Delta
V_{vp}^C\bigr|\psi\bigr>- \bigl<\psi\bigl|\frac{Z\alpha}{r}\Delta
V_{vp}^C\bigr|\psi\bigr>+\bigl<\psi\bigl|\frac{Z\alpha}{r}\bigr|\psi\bigr>~\bigl<\psi\bigl|\Delta
V_{vp}^C\bigr|\psi\bigr>.
\end{displaymath}
In matrix elements \eqref{eq:30}, the integration is performed analytically over the coordinates and then numerically
by the spectral parameter. Here are final numerical values of the matrix elements for
four parts of the Breit potential \eqref{eq:1} (relativistic term, contact term, relativistic-recoil term and spin-orbit term):
\begin{equation}
\label{eq:31}
\Delta E^{B,vp}_{sopt,~1}=\begin{cases}
^6_3Li:18.57~meV,&^7_3Li:18.79~meV\\
^9_4Be:68.57~meV,&^{10}_4Be:68.95~meV\\
^{10}_5B:184.95~meV,&^{11}_5B:185.78~meV
\end{cases},
\end{equation}
\begin{equation}
\label{eq:32}
\Delta E^{B,vp}_{sopt,~2}=\begin{cases}
^6_3Li:-8.98~meV,&^7_3Li:-9.06~meV\\
^9_4Be:-33.20~meV,&^{10}_4Be:-33.34~meV\\
^{10}_5B:-90.10~meV,&^{11}_5B:-90.42~meV
\end{cases},
\end{equation}
\begin{equation}
\label{eq:33}
\Delta E^{B,vp}_{sopt,~3}=\begin{cases}
^6_3Li:0.18~meV,&^7_3Li:0.15~meV\\
^9_4Be:0.44~meV,&^{10}_4Be:0.40~meV\\
^{10}_5B:1.08~meV,&^{11}_5B:0.98~meV
\end{cases},
\end{equation}
\begin{equation}
\label{eq:34}
\Delta E^{B,vp}_{sopt,~4}=\begin{cases}
^6_3Li:-0.89~meV,&^7_3Li:-0.90~meV\\
^9_4Be:-4.07~meV,&^{10}_4Be:-4.08~meV\\
^{10}_5B:-12.64~meV,&^{11}_5B:-12.67~meV
\end{cases}.
\end{equation}

Another corrections of the second order PT shown in Fig.\ref{fig1}(d,e,f)), have the similar structure.
They appear after the replacements $\Delta V_{vp}^C\to \Delta V^B$ and $\Delta V^C_{vp}\to \Delta
V^C_{vp,vp}$ in the basic amplitude presented in Fig.\ref{fig1}(c). Finally, the remaining two-loop corrections
in second order PT appear when one makes a replacement in Fig.\ref{fig1}(c) $\Delta V_{vp}^C\to \Delta V_{vp}^B$.
In general, the calculation of the matrix elements in this case is quite similar to expressions \eqref{eq:27}-\eqref{eq:28}.
The total value of two-loop corrections in the second order PT is included in Tables~\ref{tb1},\ref{tb2},\ref{tb3}.

\begin{figure}[htbp]
\centering
\includegraphics[scale=0.7]{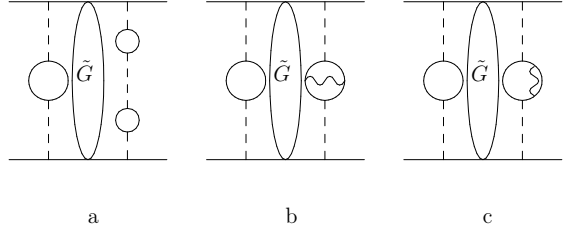}
\caption{Three-loop vacuum polarization corrections in the
second order perturbation theory. $\tilde G$ is the reduced
Coulomb Green function. Dashed line represents the Coulomb photon.}
\label{fig2}
\end{figure}

Three-loop vacuum polarization contributions in the second order PT are presented in Fig.\ref{fig2}.
The perturbation potentials which should be used in this case, are determined by
relations \eqref{eq:2}, \eqref{eq:6}, \eqref{eq:9}. Omitting the intermediate expressions,
we give only numerical values of these contributions:
\begin{equation}
\label{eq:35}
\Delta E^{vp-vp,vp}_{sopt}(2P-2S)=\begin{cases}
^6_3Li:0.04~meV,&^7_3Li:0.04~meV\\
^9_4Be:0.10~meV,&^{10}_4Be:0.10~meV\\
^{10}_5B:0.19~meV,&^{11}_5B:0.19~meV
\end{cases},
\end{equation}
\begin{equation}
\label{36}
\Delta E^{2-loop~vp,vp}_{sopt}(2P-2S)=\begin{cases}
^6_3Li:0.04~meV,&^7_3Li:0.04~meV\\
^9_4Be:0.09~meV,&^{10}_4Be:0.09~meV\\
^{10}_5B:0.16~meV,&^{11}_5B:0.16~meV
\end{cases}.
\end{equation}

In the third order of perturbation theory (topt) there exists also three-loop VP correction
of order $\alpha^3(Z\alpha)^2$ which is determined by the following relation \cite{kn2009,sgk2009}:
\begin{equation}
\label{eq:37}
\Delta E_{topt}^{3~loop~vp}=<\psi_2|\Delta V^C\tilde G\Delta V^C\tilde G\Delta V^C|\psi_2>-
<\psi_2|\Delta V^C|\psi_2><\psi_2|\Delta V^C\tilde G\tilde G\Delta V^C|\psi_2>.
\end{equation}
Its numerical contribution to the Lamb shift is equal to
\begin{equation}
\label{eq:38}
\Delta E_{topt~}^{3~loop~vp}(2P-2S)=\begin{cases}
^6_3Li:0.05~meV,&^7_3Li:0.04~meV\\
^9_4Be:0.05~meV,&^{10}_4Be:0.05~meV\\
^{10}_5B:0.31~meV,&^{11}_5B:0.31~meV
\end{cases}.
\end{equation}

\section{Nuclear structure and vacuum polarization effects}

\begin{figure}[htbp]
\centering
\includegraphics[scale=0.7]{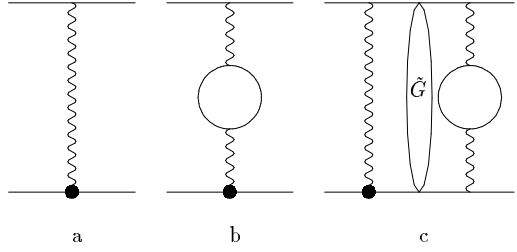}
\caption{Leading order nuclear structure and vacuum
polarization corrections. Thick point represents the nuclear vertex operator.}
\label{fig3}
\end{figure}

The second effect of the Lamb shift, comparable in magnitude to the effect of vacuum polarization,
is a nuclear structure effect. In the leading order $(Z\alpha)^4$ it is determined by the nuclear
charge radius $r_N$ after an expansion of nuclear electric form factor as follows (Fig.~\ref{fig3}(a)):
\begin{equation}
\label{eq:39}
\Delta E_{str}(2P-2S)=-\frac{\mu^3(Z\alpha)^4}{12}<r^2_N>=\begin{cases}
^6_3Li:-3674.69~meV,&^7_3Li:-3300.70~meV\\
^9_4Be:-11200.03~meV,&^{10}_4Be:-9825.73~meV\\
^{10}_5B:-25492.50~meV,&^{11}_5B:-25115.11~meV
\end{cases},
\end{equation}
where we take the nuclear charge radii from \cite{marinova} for numerical estimates.
The growth of the absolute value of the contribution \eqref{eq:39} is due to two factors $r_N$ and $Z^4$.
The signs in formulas \eqref{eq:5} and \eqref{eq:39} are opposite, thus a significant reduction of the sum
\eqref{eq:5} and \eqref{eq:39} occurs at a certain $r_N$ and $Z$. As a result this leads to significant decrease
in total value of the Lamb shift.

\begin{figure}[htbp]
\centering
\includegraphics[scale=0.7]{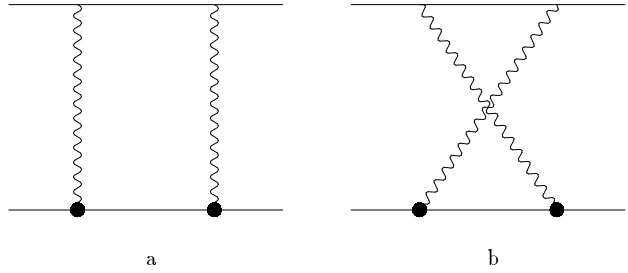}
\caption{Nuclear structure corrections of order $(Z\alpha)^5$.
Thick point is the nuclear vertex operator.}
\label{fig4}
\end{figure}

In the next to leading order $(Z\alpha)^5$ there is nuclear structure correction which is
defined by one-loop exchange diagrams (Fig.~\ref{fig4}).
Introducing only the charge form factor of the nucleus, we can represent the contribution
on finite size of nucleus to the shift of $S$-levels in the form:
\begin{equation}
\label{eq:40}
\Delta E_{str}^{2\gamma}(nS)=-\frac{\mu^3(Z\alpha)^5}{\pi n^3}\delta_{l0}
\int_0^\infty\frac{dk}{k}V(k),
\end{equation}
\begin{equation}
\label{eq:41}
V(k)=\frac{2(F^2-1)}{m_1m_2}+\frac{8m_1[-F(0)-
4m_2^2F'(0)]}{m_2(m_1+m_2)k} +\frac{k^2}{2m_1^3m_2^3}\times
\end{equation}
\begin{displaymath}
\times\left[2(F^2-1)(m_1^2+m_2^2)-F^2m_1^2\right]
+\frac{\sqrt{k^2+4m_1^2}}{2m_1^3m_2(m_1^2-m_2^2)k}\times
\end{displaymath}
\begin{displaymath}
\times\Biggl\{k^2\left[2(F^2-1)m_2^2-F^2m_1^2\right]
+8m_1^4F^2+\frac{16m_1^4m_2^2(F^2-1)}{k^2}\Biggr\}-
\end{displaymath}
\begin{displaymath}
-\frac{\sqrt{k^2+4m_2^2}m_1}{2m_2^3(m_1^2-m_2^2)k}\Biggl\{k^2
\left[2(F^2-1)-F^2\right]+8m_2^2F^2+\frac{16m_2^4(F^2-1)}{k^2}\Biggr\},
\end{displaymath}
where a subtraction of the point-like contribution and iteration term of quasipotential
is made. To perform numerical integration in \eqref{eq:40} we use dipole and Gaussian
parameterizations for the charge form factor:
\begin{equation}
\label{eq:42}
F_D(k^2)=\frac{\Lambda^4}{(k^2+\Lambda^2)^2},~\Lambda^2=\frac{12}{<r^2_N>},~~~
F_G(k^2)=e^{-\frac{1}{6}k^2r_N^2}.
\end{equation}
Numerical values of this correction for muonic atoms are the following (the result for Gaussian
parameterization is in round brackets):
\begin{equation}
\label{eq:43}
\Delta E_{str}^{2\gamma}(2P-2S)=\begin{cases}
^6_3Li:207.83~(190.63)~meV,&^7_3Li:176.67~(162.05)~meV\\
^9_4Be:826.35~(757.77)~meV,&^{10}_4Be:678.62~(622.30)~meV\\
^{10}_5B:2268.56~(2080.24)~meV,&^{11}_5B:2217.00~(2032.89)~meV
\end{cases}.
\end{equation}
We observe a significant change in the value of this contribution (approximately 10$\%$)
in the transition from the dipole to the Gaussian parameterizations.

To increase an accuracy of the Lamb Shift calculation we have to consider corrections,
which are determined by the nuclear structure effects and vacuum polarization simultaneously.
In one-photon interaction corresponding contribution is represented by the amplitude in Fig.\ref{fig3}(b).
To obtain a particle interaction operator we make an expansion of the charge form factor in momentum
representation and replace the conventional Coulomb potential to the potential of the vacuum polarization.
Then in coordinate representation we obtain:
\begin{equation}
\label{eq:44}
\Delta V^{vp}_{str}(r)=\frac{2}{3}\pi Z\alpha<r^2_N>\frac{\alpha}{3\pi}
\int_1^\infty\rho(\xi)d\xi\left[\delta({\bf r})-\frac{m_e^2\xi^2}{\pi r} e^{-2m_e\xi r}\right].
\end{equation}
Averaging \eqref{eq:44} over wave functions we find the following integral expressions for the corrections
to the levels 2S and 2P and their numerical values in the Lamb shift:
\begin{equation}
\label{eq:45}
\Delta E_{str}^{vp}(2S)=\frac{\alpha(Z\alpha)^4<r_N^2>\mu^3}{36\pi}
\int_1^\infty\rho(\xi)d\xi\frac{8a_1^3\xi^3+11a_1^2\xi^2+8a_1\xi+2}{2(a_1\xi+1)^4},~a_1=\frac{2m_e}{W},
\end{equation}
\begin{equation}
\label{eq:46}
\Delta E^{vp}_{str}(2P)=-\frac{\alpha(Z\alpha)^4\mu^3<r_N^2>}{72\pi}
\int_1^\infty\rho(\xi)d\xi\frac{a_1^2\xi^2}{(a_1\xi+1)^4},
\end{equation}
\begin{equation}
\label{eq:47}
\Delta E^{vp}_{str}(2P-2S)=\begin{cases}
^6_3Li:-14.21~meV,&^7_3Li:-12.78~meV\\
^9_4Be:-48.35~meV,&^{10}_4Be:-42.44~meV\\
^{10}_5B:-118.86~meV,&^{11}_5B:-117.14~meV
\end{cases}.
\end{equation}

The same order $\alpha(Z\alpha)^4$ contribution is given by the amplitude in the
second order PT presented in Fig.\ref{fig3}(c):
\begin{equation}
\label{eq:48}
\Delta E^{vp}_{str,sopt}(2P-2S)=-\frac{\alpha(Z\alpha)^4\mu^3<r_N^2>}
{36\pi}\int_1^\infty\rho(\xi)d\xi\times
\end{equation}
\begin{displaymath}
\times\frac{4(a_1\xi+1)(2a_1^2\xi^2+1)\ln(a_1\xi+1)+a_1\xi(4a_1\xi(a_1\xi(a_1\xi+3)+1)+11)+3}
{(a_1\xi+1)^5}=
\end{displaymath}
\begin{displaymath}
=\begin{cases}
^6_3Li:-23.00~meV,&^7_3Li:-20.68~meV\\
^9_4Be:-80.52~meV,&^{10}_4Be:-70.68~meV\\
^{10}_5B:-202.98~meV,&^{11}_5B:-200.06~meV
\end{cases}.
\end{displaymath}
\begin{figure}[htbp]
\centering
\includegraphics[scale=0.7]{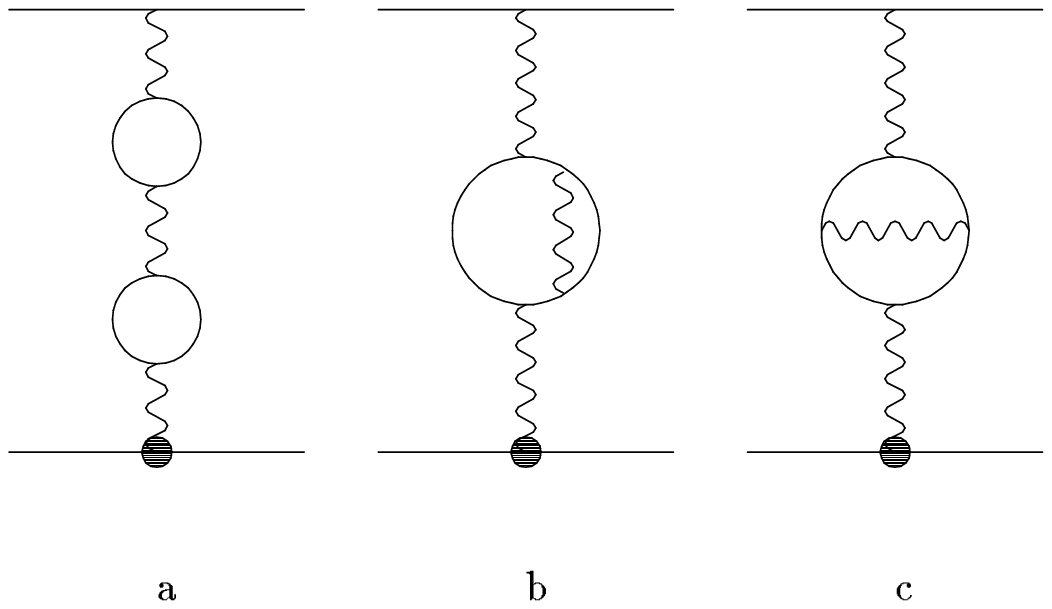}
\caption{Nuclear structure and two-loop vacuum polarization
effects in the one-photon interaction. Thick point is the
nuclear vertex operator.}
\label{fig5}
\end{figure}

Bearing in mind that the quantities \eqref{eq:47} and \eqref{eq:48} are large
we evaluate also nuclear structure corrections with the account of two-loop
vacuum polarization effects in $1\gamma$-interaction (Fig.~\ref{fig5}(a-c)).
The method of constructing the potentials is the same as in this and preceding sections.
Corresponding potentials have the following form:
\begin{equation}
\label{eq:49}
\Delta V^{vp-vp}_{str}(r)=\frac{2}{3}Z\alpha<r_N^2>\left(\frac{\alpha}
{3\pi}\right)^2\int_1^\infty\rho(\xi)d\xi\int_1^\infty\rho(\eta)d\eta\times
\end{equation}
\begin{displaymath}
\times\left[\pi\delta({\bf r})-\frac{m_e^2}{r(\xi^2-\eta^2)}\left
(\xi^4 e^{-2m_e\xi r}-\eta^4e^{-2m_e\eta r}\right)\right],
\end{displaymath}
\begin{equation}
\label{eq:50}
\Delta V^{2-loop~vp}_{str}(r)=\frac{4}{9}Z\alpha<r_N^2>\left
(\frac{\alpha}{\pi}\right)^2\int_0^1\frac{f(v)dv}{1-v^2}\left[\pi\delta({\bf
r})- \frac{m_e^2}{r(1-v^2)}e^{-\frac{2m_er}{\sqrt{1-v^2}}}\right].
\end{equation}
The sum of corrections to the Lamb shift $(2P-2S)$ that are provided by \eqref{eq:49}
and \eqref{eq:50} is equal
\begin{equation}
\label{eq:51}
\Delta E_{str}^{vp,vp}(2P-2S)=\begin{cases}
^6_3Li:-0.22~meV,&^7_3Li:-0.20~meV\\
^9_4Be:-0.82~meV,&^{10}_4Be:-0.70~meV\\
^{10}_5B:-2.15~meV,&^{11}_5B:-2.12~meV
\end{cases}.
\end{equation}

\begin{figure}[h]
\centering
\includegraphics[scale=0.8]{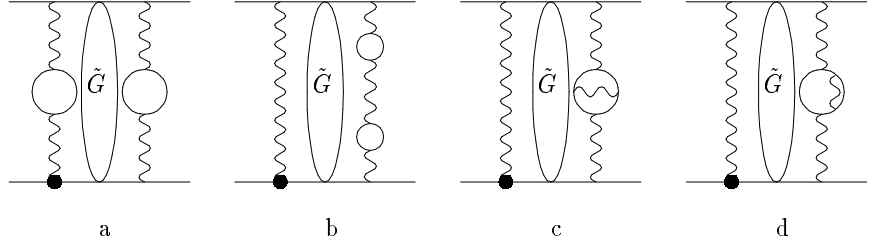}
\caption{Nuclear structure and two-loop vacuum polarization
effects in the second order perturbation theory. Thick point is
the nuclear vertex operator. $\tilde G$ is the reduced Coulomb
Green function.}
\label{fig6}
\end{figure}

\begin{figure}[htbp]
\centering
\includegraphics[scale=0.8]{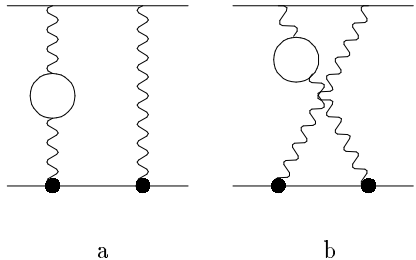}
\caption{Nuclear structure and electron vacuum polarization
effects in the two-photon exchange diagrams. Thick point is
the nuclear vertex operator.}
\label{fig7}
\end{figure}

There are two-loop VP corrections with nuclear structure of order $\alpha^2(Z\alpha)^4$
in the second order PT (see Fig.~\ref{fig6}(a-d)). They can be calculated as \eqref{eq:45}
with the replacement one-loop VP potential to two-loop. Their numerical values are included in
Table~\ref{tb1},\ref{tb2},\ref{tb3}. An important role plays a correction by two-photon exchange
diagrams with the effect of vacuum polarization, as it reinforced by the factor $Z^5$ (see Fig.~\ref{fig7}).
An analytical expression for this correction and its numerical value is defined by modified
potential $V(p)$ from \eqref{eq:41}:
\begin{equation}
\label{eq:52}
\Delta E^{2\gamma}_{str,vp}(nS)=-\frac{2\mu^3\alpha(
Z\alpha)^5}{\pi^2 n^3} \int_0^\infty k V(k) dk
\int_0^1\frac{v^2(1-\frac{v^2}{3})dv}{k^2(1-v^2)+4m_e^2},
\end{equation}
\begin{equation}
\label{eq:53}
\Delta E_{str,vp}^{2\gamma}(2P-2S)=\begin{cases}
^6_3Li:3.19 (3.05)~meV,&^7_3Li:2.73 (2.61)~meV\\
^9_4Be:12.74 (12.08)~meV,&^{10}_4Be:10.80 (10.05)~meV\\
^{10}_5B:35.07 (33.37)~meV,&^{11}_5B:33.67 (32.63)~meV
\end{cases}.
\end{equation}

Since expression \eqref{eq:52} contains the charge form factor of the nucleus, here we present two numerical values
of the contribution corresponding to the parameterizations in \eqref{eq:42}.
Corrections to the Lamb shift discussed in this and previous sections are such that analytical expressions
for them are quite bulky, since the characteristic parameter $W/m_e$ is large. It can not be used
as an expansion parameter. For this reason, it is more convenient to present corrections in integral form,
what we do in this paper.

\section{Recoil corrections, muon self-energy and vacuum polarization
effects}

There is another group of corrections, which were obtained in analytical form in the study
of the Lamb shift $(2P-2S)$ in the hydrogen atom during many years. Their calculation is discussed
in detail in \cite{EGS}. Corresponding analytical results can be used directly for numerical estimates in
muonic atoms. For the sake of completeness we present in this section the key expressions for such corrections,
which have necessary order in $\alpha$ and the ratio of particle masses and provide significant numerical values
in the Lamb shift $(2P-2S)$.

It is known an analytical expression for the recoil correction of order $\alpha^4$, which was obtained
after calculating the matrix elements of the Breit potential \cite{EGS,udjpra2011,SY}:
\begin{equation}
\label{eq:54}
\Delta E_{rec}^{(Z\alpha)^4}(2P-2S)=
\begin{cases}
\frac{\mu^3(Z\alpha)^4}{48m_2^2},~~~\delta_I=1,\\
\frac{\mu^3(Z\alpha)^4}{12m_2^2},~~~\delta_I=0,\\
\end{cases}
=
\begin{cases}
^6_3Li: 0.68~meV,&^7_3Li:0.13~meV\\
^9_4Be: 0.24~meV,&^{10}_4Be:0.79~meV\\
^{10}_5B:1.94~meV,&^{11}_5B:0.40~meV
\end{cases}.
\end{equation}

The recoil correction of order $(Z\alpha)^5$ is related with two-photon exchange amplitudes
in which the nucleus is considered as a point particle \cite{SY,EGS}:
\begin{equation}
\label{eq:55}
\Delta E_{rec}^{(Z\alpha)^5}=\frac{\mu^3(Z\alpha)^5}{m_1m_2\pi
n^3}\bigl[(\frac{2}{3}\ln\frac{1}{Z\alpha}-\frac{1}{9})\delta_{l0}
-\frac{8}{3}\ln k_0(n,l)-
\frac{7}{3}a_n-\frac{2}{m_2^2-m_1^2}\delta_{l0}(m_2^2\ln\frac{m_1}{\mu}-
m_1^2\ln\frac{m_2}{\mu})\bigr],
\end{equation}
where $\ln k_0(n,l)$ is the Bethe logarithm \cite{EGS,drake1990}:
\begin{equation}
\label{eq:56}
\ln k_0(2S)=2.811769893120563,
\end{equation}
\begin{equation}
\label{eq:57}
\ln k_0(2P)=-0.030016708630213,
\end{equation}
\begin{equation}
\label{eq:58}
a_n=-2\left[\ln\frac{2}{n}+(1+\frac{1}{2}+...+\frac{1}{n}+1-\frac{1}{2n}\right]\delta_{l0}+
\frac{(1-\delta_{l0})}{l(l+1)(2l+1)}.
\end{equation}
The expression \eqref{eq:55} gives the following numerical result:
\begin{equation}
\label{eq:59}
\Delta E_{rec}^{(Z\alpha)^5}(2P-2S)=
\begin{cases}
^6_3Li: -2.15~meV,&^7_3Li:-1.86~meV\\
^9_4Be: -5.97~meV,&^{10}_4Be:-5.40~meV\\
^{10}_5B:-16.03~meV,&^{11}_5B:-14.63~meV
\end{cases}.
\end{equation}
Numerical result for recoil correction of order $(Z\alpha)^6$ is presented in Tables~\ref{tb1},\ref{tb2},\ref{tb3}
according to analytical formula from \cite{EGS,EG}.

It should be noted a significant value contribution, which is given by the radiative corrections to the muon
line, corrections from the Dirac and Pauli form factors of the muon and muon vacuum polarization (mvp).
It is appropriate to quote here the relevant analytical formulas \cite{EG1,EGS}:
\begin{equation}
\label{eq:60}
\Delta E_{mvp,mse}(2S)=\frac{\alpha(Z\alpha)^4}{8\pi}\frac{\mu^3}{m_1^2}
\Biggl[\frac{4}{3}\ln\frac{m_1}{\mu(Z\alpha)^2}-\frac{4}{3}\ln
k_0(2S)+ \frac{38}{45}+
\end{equation}
\begin{displaymath}
+\frac{\alpha}{\pi}\left(-\frac{9}{4}\zeta(3)+\frac{3}{2} \pi^2\ln
2-\frac{10}{27}\pi^2-\frac{2179}{648}\right)+4\pi Z\alpha\left(
\frac{427}{384}-\frac{\ln 2}{2}\right)\Biggr],
\end{displaymath}
\begin{equation}
\label{eq:61}
\Delta E_{mvp,mse}(2P)=\frac{\alpha(Z\alpha)^4}{8\pi}\frac{\mu^3}{m_1^2}
\Biggl[-\frac{4}{3}\ln k_0(2P)-\frac{m_1}{6\mu}-
\end{equation}
\begin{displaymath}
-\frac{\alpha}{3\pi}\frac{m_1}{\mu}\left(\frac{3}{4}\zeta(3)-\frac{
\pi^2}{2}\ln 2+\frac{\pi^2}{12}+\frac{197}{144}\right)\Biggr],
\end{displaymath}
which lead to numerical results:
\begin{equation}
\label{eq:62}
\Delta E_{mse,mvp}(2P-2S)=
\begin{cases}
^6_3Li: -50.99~meV,&^7_3Li:-51.36~meV\\
^9_4Be: -149.00~meV,&^{10}_4Be:-149.52~meV\\
^{10}_5B:-337.45~meV,&^{11}_5B:-338.40~meV
\end{cases}.
\end{equation}

Significantly smaller size are the radiative-recoil corrections of orders
$\alpha(Z\alpha)^5$ and $(Z^2\alpha)(Z\alpha)^4$ from the
Tables 8-9 \cite{EGS} (see their explicit form in \cite{jetp2015}).
Their numerical values we have included in the summary Tables~\ref{tb1}-\ref{tb3}.

On the basis of obtained in \cite{Friar1,LYE} expressions we give an estimate of
the nuclear structure corrections of orders $(Z\alpha)^6$ and $\alpha(Z\alpha)^5$
to the lamb shift in muonic ions. So, for the structure correction of order $(Z\alpha)^6$
we obtain:
\begin{equation}
\label{eq:63}
\Delta E_{str}^{(Z\alpha)^6}(2P-2S)=\frac{(Z\alpha)^6}{12}\mu^3\Bigl\{r_N^2
\left[\langle\ln \mu Z\alpha
r\rangle+C-\frac{3}{2}\right]-\frac{1}{2}r_N^2+
\frac{1}{3}\langle r^3\rangle\langle\frac{1}{r}\rangle-
\end{equation}
\begin{displaymath}
-I_2^{rel}-I_3^{rel} -\mu^2F_{NR}+\frac{1}{40}\mu^2\langle
r^4\rangle\Bigr\} =
\begin{cases}
^6_3Li: -7.38~meV,&^7_3Li:-6.69~meV\\
^9_4Be: -40.60~meV,&^{10}_4Be:-35.70~meV\\
^{10}_5B:-145.27~meV,&^{11}_5B:-143.04~meV
\end{cases}.
\end{displaymath}
where the quantities $I_{2,3}^{rel}, F_{NR}$ are written explicitly in \cite{Friar1}.
Significant growth of the numerical values in \eqref{eq:63} in the transition from one to the other muonic
ions muon is caused by two factors $Z$ and $r_N$. Summary result for corrections of orders
$(Z\alpha)^6$ and $\alpha(Z\alpha)^5$ is presented in Tables~\ref{tb1},\ref{tb2},\ref{tb3}.

\begin{figure}[htbp]
\centering
\includegraphics[scale=0.8]{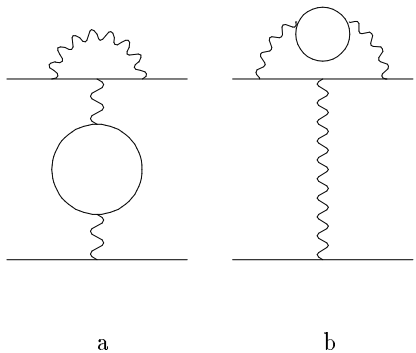}
\caption{Radiative corrections with the vacuum polarization effects.}
\label{fig8}
\end{figure}

An amplitude in Fig.\ref{fig8}(b) gives the contribution to the
energy spectrum, which can be expressed in terms of the slope of
the Dirac form factor $F_1'$ and the Pauli form factor $F_2$.
Numerical values in the Lamb shift are obtained by means of two-loop corrections
to form factors $F_1'(0)$ and $F_2(0)$ which were calculated in \cite{BCR}.
Another contribution with vacuum polarization in Fig.~\ref{fig8}(a)
was investigated in \cite{KP1,jw}. It is included in final Tables on separate line.
The contribution of hadron vacuum polarization to the Lamb shift can be derived by means of corresponding
result for muonic hydrogen \cite{borie1981,Friar2,M6}.

\section{Summary and conclusion}

In this study, we perform a calculation of the Lamb shift $(2P_{1/2}-2S_{1/2})$ in the number of muonic ions with
different nuclear charge and the nuclear charge radius. Different corrections with fairly high degrees of fine
structure  constant $\alpha^3\div\alpha^6$ have been taken into account.
All contributions that are analyzed may be divided into two groups.
The first group includes the corrections specific to each muonic ion, which are presented in the integral form
and calculated analytically and numerically. The second group of corrections is obtained on the basis of the known
analytical expressions derived in the study of the Lamb shift in the hydrogen atom.
Numerical values of all corrections are written explicitly in Tables~\ref{tb1}-\ref{tb3}.
The resulting total numerical values of shifts in the muonic ions of lithium, beryllium and boron
can be used for comparison with future experimental data.
These numerical values allow us to trace the dynamics of changes in the values of corrections during
the transition from one ion to another.

It is known that the position of the energy levels of the $2S_{1/2}$ and $2P_{1/2}$ in atoms of electronic
hydrogen and muonic hydrogen differs significantly. A similar change we have seen in the study of muonic ions $(\mu Li)^{2+}$,
$(\mu Be)^{3+}$ and $(\mu B)^{4+}$:
if a muonic lithium ion $2P$ level is above the $2S$ level, then for ions of muonic beryllium and boron
we get the reverse arrangement of levels. This effect is due to compensate for two basic contributions
to the Lamb shift from a one-loop electronic vacuum polarization and nuclear structure of order $(Z\alpha)^4$.
As a result the corrections of higher order in $\alpha$, enhanced by nuclear charge degrees, become more important.

As noted above, the problem of the Lamb shift in muonic ions of lithium, beryllium and boron
was studied many years ago in \cite{Drake1}. One part of results in \cite{Drake1} was obtained with the use
of nonrelativistic wave functions and treating the finite nuclear size and vacuum-polarization potentials
as small perturbations. It is consistent with our results within a small change in the fundamental physical constants.
Another part of the results in \cite{Drake1} was obtained by means of numerical solution of the Dirac equation
and treating the remaining small corrections due to the muon self-energy, the higher-order K\"allen-Sabrey vacuum
polarization term, and nuclear polarization by perturbation theory. To compare our results with such calculation in
\cite{Drake1} we must take the sum of a few lines of our Tables, corresponding to the first and second-order
perturbation theory, and in the case of effects on the structure of the nucleus to the sum of corrections of one-photon
and two-photon amplitudes. Since our other corrections are numerically small, it is convenient further to compare
the complete results for the Lamb shift. Our values of the charge radii of nuclei are slightly different
from the values of \cite{Drake1}, what is one of the reasons for the differences of total numerical results.
Another reason is related to corrections of a higher order, accounted for in this paper.
For example, we use the value of the charge radius of lithium nucleus $r(_3^6Li)=2.5890$ fm from \cite{marinova}.
It is slightly different from the value 2.560 fm used in \cite{Drake1}. This difference gives an addition of 82 meV to
our result from Table~\ref{tb1} due to structure correction of order $(Z\alpha)^4$ \eqref{eq:39}. In turn, our structure
correction of order $(Z\alpha)^5$ \eqref{eq:43} becomes smaller on 7 meV and reduces essentially the divergence
from the value of the Lamb shift in \cite{Drake1}. It is necessary also to point out that the nuclear charge density
distribution in our work and in \cite{Drake1} are slightly different. The same situation occurs for the other nuclei.

Note also that there is another contribution to the polarizability of the nucleus, for which we use an estimate from \cite{Drake1}.
It is expressed in terms of the (-2) moment of the total electric-dipole photo-absorption cross section $\sigma_{-2}$.
In the case of light nuclei there exists a simple formula $\sigma_{-2}=3.5\cdot k\cdot A^{5/3}~\frac{\mu b}{MeV}$ \cite{orce} which
allows to obtain an estimate for the cross sections of lithium, beryllium and boron. The main uncertainty of the calculation
is related with nuclear structure and polarizability corrections. The errors of corresponding contributions to the
Lamb shift are presented explicitly in Tables, where we have indicated only one total value of the Lamb shift corresponding to
dipole parameterization for the nuclear charge form factor. The uncertainties connected with the value of nuclear charge radius
appear at the calculation of other structure corrections but they are much less basic.
Weak interaction contribution is very small and is not considered in this study (see \cite{eides}).
Thus, with proper experimental accuracy we can obtain more precise values of the nuclear charge radii.

\begin{acknowledgments}
This work is supported by the Russian Foundation for Basic Research (grant No. 16-02-00554) and
Ministry of Education and Science of Russia (grant No. 1394).
\end{acknowledgments}

\newpage

\begin{table}[htbp]
\caption{\label{tb1}Lamb shift $(2P_{1/2}-2S_{1/2})$ in muonic
ions $(\mu~^ 7_3Li)^{2+}$ and $(\mu~^ 6_3Li)^{2+}$.}
\bigskip
\begin{ruledtabular}
\begin{tabular}{|c|c|c|}  \hline
Contribution to the splitting &$(\mu~^ 7_3Li)^{2+}$,~meV  &$(\mu~^ 6_3Li)^{2+}$, meV  \\
\hline 1&2&3 \\
\hline
VP contribution of order $\alpha(Z\alpha)^2$ & 4682.38 &4664.95   \\
in one-photon interaction &  &   \\   \hline
Two-loop VP contribution &32.54    & 32.41  \\
of order $\alpha^2(Z\alpha)^2$   &   &   \\    \hline
VP and MVP contribution in  &0.01     &0.01      \\
one-photon interaction &     &     \\    \hline
Three-loop VP contribution in  &0.17    & 0.17    \\
one-photon interaction &    &     \\    \hline
The Wichmann-Kroll
correction  & -0.09  &  -0.09  \\    \hline
Light-by-Light scattering
correction&0.03&0.03\\    \hline
Relativistic and VP corrections of order  & -6.13   & -6.07     \\
$\alpha(Z\alpha)^4$ in the first order PT      &       &    \\    \hline
Relativistic and two-loop VP  & -0.02    &  -0.02    \\
corrections of order $\alpha^2(Z\alpha)^4$   &       &    \\
in the first order PT   &       &    \\   \hline
Two-loop VP contribution of order   &5.66  &   5.63  \\
$\alpha^2(Z\alpha)^2$ in the second order PT &    &    \\   \hline
Two-loop MVP and EVP contribution &0.02& 0.02\\
in the second order PT& & \\ \hline
Relativistic and one-loop VP  & 8.98    &   8.88  \\
corrections of order $\alpha(Z\alpha)^4$  &       &    \\
in the second order PT  &    &    \\    \hline
Three-loop VP contribution in the           &0.08  & 0.08  \\
second order PT of order $\alpha^3(Z\alpha)^2$  &      &    \\   \hline
Three-loop VP contribution & 0.04& 0.05 \\
in the third order PT of order $\alpha^3(Z\alpha)^2$& &\\   \hline
Relativistic and two-loop VP   & 0.15    & 0.15    \\
corrections of order $\alpha^2(Z\alpha)^4$  &       &    \\
in the second order PT  &    &    \\    \hline
Nuclear structure
contribution of order $(Z\alpha)^4$   &$-3301\pm 117$  &$-3675\pm 112$ \\  \hline
Nuclear structure contribution   & $177\pm 9$ & $208\pm 9$   \\
of order $(Z\alpha)^5$ from $2\gamma$ amplitudes
&          &    \\   \hline
Nuclear structure and VP contribution  &-12.78 &-14.21    \\
in $1\gamma$ interaction of order $\alpha(Z\alpha)^4$ &   &  \\    \hline
Nuclear structure and VP contribution   &-20.68 & -23.00   \\
in the second order PT of order $\alpha(Z\alpha)^4$ &   &  \\    \hline
\end{tabular}
\end{ruledtabular}
\end{table}
\begin{table}[htbp]
Table I (continued).\\
\bigskip
\begin{ruledtabular}
\begin{tabular}{|c|c|c|}  \hline
1&2&3 \\     \hline
Nuclear structure and two-loop VP  &-0.11  & -0.11   \\
contribution in $1\gamma$ interaction of order
$\alpha^2(Z\alpha)^4$ &   &  \\
\hline
Nuclear structure and two-loop VP contribution&-0.30 &  -0.33  \\
in the second order PT of order $\alpha^2(Z\alpha)^4$ &   &  \\
\hline
Nuclear structure contribution of order   &2.73 & 3.19  \\
$\alpha(Z\alpha)^5$ from $2\gamma$ amplitudes with VP insertion &
&    \\
\hline Recoil correction of order $(Z\alpha)^4$ &0.13&  0.68 \\
\hline Recoil correction of order $(Z\alpha)^5$ &-1.86&-2.15   \\
\hline
Recoil correction of order $(Z\alpha)^6$   &0.02   &0.03   \\
\hline Muon self-energy and MVP contribution &-51.36 &-50.99 \\
\hline
Radiative-recoil corrections  &-0.12 & -0.16  \\
of orders $(Z^2\alpha)(Z\alpha)^4$  &   &
\\  \hline
Nuclear structure corrections  & -5.52 & -6.07 \\
of orders $(Z\alpha)^6$, $\alpha(Z\alpha)^5$ &  &  \\  \hline
Muon form factor $F_1'(0)$, $F_2(0)$ contributions & -0.16&-0.16  \\    \hline
Muon self-energy and VP contribution &-0.23 & -0.23\\   \hline
HVP contribution & 1.17 & 1.16\\   \hline
Nuclear polarizability  & $21\pm 4$  & $15\pm 4$   \\     \hline
Total contribution  &1531.75   &  1161.85\\
\hline
\end{tabular}
\end{ruledtabular}
\end{table}

\newpage
\begin{table}[htbp]
\caption{\label{tb2}Lamb shift $(2P_{1/2}-2S_{1/2})$ in muonic
ions $(\mu~^ {10}_4Be)^{3+}$ and $(\mu~^ {9}_4Be)^{3+}$.}
\bigskip
\begin{ruledtabular}
\begin{tabular}{|c|c|c|}  \hline
Contribution to the splitting &$(\mu~^ {10}_4Be)^{3+}$,~meV  &$(\mu~^ {9}_4Be)^{3+}$, meV  \\
\hline 1&2&3 \\     \hline
VP contribution of order $\alpha(Z\alpha)^2$ & 9270.74 &9255.79   \\
in one-photon interaction &  &   \\   \hline
Two-loop VP contribution &65.75    & 65.66   \\
of order $\alpha^2(Z\alpha)^2$   &   &   \\    \hline
VP and MVP contribution in  &0.05     &0.05      \\
one-photon interaction &     &     \\    \hline
Three-loop VP contribution in  &0.42    & 0.42    \\
one-photon interaction &    &     \\    \hline
The Wichmann-Kroll
correction  & -0.24  &  -0.24  \\    \hline
Light-by-Light scattering   correction&0.07&0.07\\      \hline
Relativistic and VP corrections of order  & -22.34    & -22.23     \\
$\alpha(Z\alpha)^4$ in the first order PT      &       &    \\
\hline
Relativistic and two-loop VP  & -0.07    &  -0.07    \\
corrections of order $\alpha^2(Z\alpha)^4$   &       &    \\
in the first order PT   &       &    \\
\hline
Two-loop VP contribution of order   &12.73   &   12.70  \\
$\alpha^2(Z\alpha)^2$ in the second order PT &    &    \\    \hline
Two-loop MVP and EVP contribution &0.04& 0.04\\
in the second order PT& & \\ \hline
Relativistic and one-loop VP  & 31.93    &    31.74  \\
corrections of order $\alpha(Z\alpha)^4$  &       &    \\
in the second order PT  &    &    \\    \hline
Three-loop VP contribution in the           &0.19  &0.19  \\
second order PT of order $\alpha^3(Z\alpha)^2$  &      &    \\    \hline
Three-loop VP contribution & 0.05&0.05 \\
in the third order PT of order $\alpha^3(Z\alpha)^2$& &\\    \hline
Relativistic and two-loop VP   & 0.55    & 0.54     \\
corrections of order $\alpha^2(Z\alpha)^4$  &       &    \\
in the second order PT  &    &    \\    \hline
Nuclear structure
contribution of order $(Z\alpha)^4$   &$-9826\pm 142$  &$-11200\pm 107$ \\     \hline
Nuclear structure contribution   & $679\pm 14$ & $826\pm 12$   \\
of order $(Z\alpha)^5$ from $2\gamma$ amplitudes   &          &    \\   \hline
Nuclear structure and VP contribution  &-42.44 &-48.35    \\
in $1\gamma$ interaction of order $\alpha(Z\alpha)^4$ &   &  \\    \hline
Nuclear structure and VP contribution   &-70.68 & -80.52   \\
in the second order PT of order $\alpha(Z\alpha)^4$ &   &  \\   \hline
\end{tabular}
\end{ruledtabular}
\end{table}

\begin{table}[htbp]
Table II (continued).\\
\bigskip
\begin{ruledtabular}
\begin{tabular}{|c|c|c|}  \hline
1&2&3 \\       \hline
Nuclear structure and two-loop VP  &-0.36  & -0.41   \\
contribution in $1\gamma$ interaction of order
$\alpha^2(Z\alpha)^4$ &   &  \\       \hline
Nuclear structure and two-loop VP contribution&-1.09  &  -1.25  \\
in the second order PT of order $\alpha^2(Z\alpha)^4$ &   &  \\    \hline
Nuclear structure contribution of order   &10.80  & 12.74   \\
$\alpha(Z\alpha)^5$ from $2\gamma$ amplitudes with VP insertion &     &    \\     \hline
Nuclear structure and three-loop VP contribution&-0.01  &  -0.01  \\
in the second order PT of order $\alpha^3(Z\alpha)^4$ &   &  \\   \hline
Recoil correction of order $(Z\alpha)^4$ &0.79& 0.24  \\   \hline
Recoil correction of order $(Z\alpha)^5$ &-5.40&-5.97   \\    \hline
Recoil correction of order $(Z\alpha)^6$   &0.10   &0.11   \\    \hline
Muon self-energy and MVP contribution &-149.52 &-149.00 \\    \hline
Radiative-recoil corrections  &-0.31 & -0.39  \\
of orders $(Z^2\alpha)(Z\alpha)^4$  &   &    \\  \hline
Nuclear structure corrections  & -31.44 & -35.44  \\
of orders $(Z\alpha)^6$, $\alpha(Z\alpha)^5$ &  &  \\     \hline
Muon form factor $F_1'(0)$, $F_2(0)$ contributions & -0.52&-0.51  \\   \hline
Muon self-energy and VP contribution &-0.71 & -0.71\\     \hline
HVP contribution & 3.75 & 3.74\\     \hline
Nuclear polarizability  & $104\pm 21$  & $82\pm 16$   \\     \hline
Total contribution  & 29.83  &-1253.02  \\
\hline
\end{tabular}
\end{ruledtabular}
\end{table}

\newpage
\begin{table}[htbp]
\caption{\label{tb3}Lamb shift $(2P_{1/2}-2S_{1/2})$ in muonic
ions $(\mu~^ {11}_5B)^{4+}$ and $(\mu~^ {10}_5B)^{4+}$.}
\bigskip
\begin{ruledtabular}
\begin{tabular}{|c|c|c|}  \hline
Contribution to the splitting &$(\mu~^ {11}_5B)^{4+}$,~meV  &$(\mu~^ {10}_5B)^{4+}$, meV  \\
\hline 1&2&3 \\    \hline
VP contribution of order $\alpha(Z\alpha)^2$ & 15375.55 &15356.42   \\
in one-photon interaction &  &   \\   \hline
Two-loop VP contribution &112.09    & 111.94   \\
of order $\alpha^2(Z\alpha)^2$   &   &   \\    \hline
VP and MVP contribution in  &0.13     &0.13     \\
one-photon interaction &     &     \\    \hline
Three-loop VP contribution in  &0.77    & 0.73    \\
one-photon interaction &    &     \\    \hline
The Wichmann-Kroll
correction  & -0.50  &  -0.50 \\
\hline
Light-by-Light scattering
correction&0.16&0.16\\
\hline
Relativistic and VP corrections of order  & -59.73   & -59.46     \\
$\alpha(Z\alpha)^4$ in the first order PT      &       &    \\
\hline
Relativistic and two-loop VP  & -0.17    &  -0.17    \\
corrections of order $\alpha^2(Z\alpha)^4$   &       &    \\
in the first order PT   &       &    \\
\hline
Two-loop VP contribution of order   &23.43  &   23.39  \\
$\alpha^2(Z\alpha)^2$ in the second order PT &    &    \\
\hline Two-loop MVP and EVP contribution &0.10& 0.10\\
in the second order PT& & \\ \hline
Relativistic and one-loop VP  & 83.67    &    83.29 \\
corrections of order $\alpha(Z\alpha)^4$  &       &    \\
in the second order PT  &    &    \\    \hline
Three-loop VP contribution in the           &0.35  &0.35  \\
second order PT of order $\alpha^3(Z\alpha)^2$  &      &    \\
\hline
Three-loop VP contribution & 0.31&0.31 \\
in the third order PT of order $\alpha^3(Z\alpha)^2$& &\\
\hline
Relativistic and two-loop VP   & 1.44   &   1.44   \\
corrections of order $\alpha^2(Z\alpha)^4$  &       &    \\
in the second order PT  &    &    \\     \hline
Nuclear structure
contribution of order $(Z\alpha)^4$   &$-25115 \pm 618$ &$-25493\pm 1059$ \\   \hline
Nuclear structure contribution   & $2217\pm 82$ & $2269\pm 143$   \\
of order $(Z\alpha)^5$ from $2\gamma$ amplitudes &          &    \\   \hline
Nuclear structure and VP contribution  &-117.14 &-118.86    \\
in $1\gamma$ interaction of order $\alpha(Z\alpha)^4$ &   &  \\  \hline
Nuclear structure and VP contribution   &-200.06 & -202.98   \\
in the second order PT of order $\alpha(Z\alpha)^4$ &   &  \\    \hline
\end{tabular}
\end{ruledtabular}
\end{table}

\begin{table}[htbp]
Table III (continued).\\
\bigskip
\begin{ruledtabular}
\begin{tabular}{|c|c|c|}  \hline
1&2&3 \\    \hline
Nuclear structure and two-loop VP  &-1.00 & -1.02   \\
contribution in $1\gamma$ interaction of order
$\alpha^2(Z\alpha)^4$ &   &  \\     \hline
Nuclear structure and two-loop VP contribution&-3.25  &  -3.30  \\
in the second order PT of order $\alpha^2(Z\alpha)^4$ &   &  \\
\hline
Nuclear structure contribution of order   &33.67  & 35.07   \\
$\alpha(Z\alpha)^5$ from $2\gamma$ amplitudes with VP insertion &  &    \\    \hline
Nuclear structure and three-loop VP contribution&-0.05  &  -0.07  \\
in the second order PT of order $\alpha^3(Z\alpha)^4$ &   &  \\   \hline
Recoil correction of order $(Z\alpha)^4$ &0.40& 1.94  \\   \hline
Recoil correction of order $(Z\alpha)^5$ &-14.63&-16.03   \\   \hline
Recoil correction of order $(Z\alpha)^6$   &0.34   &0.37  \\
\hline Muon self-energy and MVP contribution &-338.40 &-337.45 \\   \hline
Radiative-recoil corrections  &-0.91 & -1.10  \\
of orders $(Z^2\alpha)(Z\alpha)^4$  &   &    \\  \hline
Nuclear structure corrections  & -128.14 & -130.15  \\
of orders $(Z\alpha)^6$, $\alpha(Z\alpha)^5$ &  &  \\   \hline
Muon form factor $F_1'(0)$, $F_2(0)$ contributions & -1.26&-1.26  \\    \hline
Muon self-energy and VP contribution &-1.66 & -1.66\\     \hline
HVP contribution & 9.18 & 9.15\\    \hline
Nuclear polarizability  & $122\pm 24$  & $103\pm 21$   \\     \hline
Total contribution  &   -8001.31&-8370.22  \\
\hline
\end{tabular}
\end{ruledtabular}
\end{table}

\end{document}